\documentclass[aps,prl,twocolumn,groupedaddress,showpacs]{revtex4-1}
\usepackage{amsmath}
\usepackage{relsize}
\usepackage{graphicx}
\usepackage{textcomp}

\newcommand {\beq} {\begin{eqnarray}}
\newcommand {\eeqn} [1] {\label{#1} \end{eqnarray}}%

\begin{document}

\title{Reply to Comment on "Sensitivity of $(d,p)$ reactions to high $n$-$p$ momenta and the consequences for nuclear spectroscopy studies"
 }
\author { G.W. Bailey, N.K. Timofeyuk and J.A. Tostevin}
\address{Department of Physics, University of Surrey, Guildford, Surrey GU2 7XH, U.K.}
\maketitle

\vskip 0.3 cm

We point out that after presenting our results on high $n$-$p$ momentum sensitivity of the $(d,p)$ cross sections in \cite{Bai16}
the last paragraph of our Letter refers to a need of going beyond the leading order of Weinberg state treatment. This task could be achieved by using any method that can provide exact solution of the three-body problem.    Deltuva  uses Faddeev equations to study the NN-model dependence of the $(d,p)$ cross sections \cite{Del18}. His results are consistent with a new study performed at Surrey which is undergoing a reviewing process at Physical Review C. Both studies discuss the $n$-$p$ sensitivity   within three-body $n+p+A$ models with $NN$-independent $N$-$A$ optical potentials. The sensitivity may   reappear in many-body treatment of $(d,p)$ reactions, for example,  due to the threshold position dependence.


\begin{thebibliography}{}
\bibitem{Bai16} G.W. Bailey, N.K. Timofeyuk and J.A. Tostevin, Phys. Rev. Lett. 117 162502 (2016)
\bibitem{Del18} A. Deltuva,  arXiv:1806.00298 (2018)
\end{thebibliography}
\end{document}